\documentclass[]{spie} 
\usepackage[]{graphicx}
\usepackage{amsmath}
\usepackage{listings}
\usepackage{url}

\title{The 2012 Interferometric Imaging Beauty Contest} 

\author{Fabien Baron\supit{a}, William D. Cotton\supit{b},
Peter~R.~Lawson\supit{c}, Steve~T.~Ridgway\supit{d}, Alicia Aarnio\supit{a},
John D. Monnier\supit{a}, Karl-Heinz Hofmann\supit{e}, Dieter Schertl\supit{e},
Gerd Weigelt\supit{e}, \'Eric Thi\'ebaut\supit{f}, F\'err\'eol Soulez\supit{f},
David Mary\supit{g}, Florentin Millour\supit{g}, Martin Vannier\supit{g}, John
Young\supit{h}, Nicholas M. Elias II \supit{i}, Henrique R. Schmitt \supit{i},
Sridharan Rengaswamy\supit{j}  
\skiplinehalf
\supit{a}Univ. of Michigan, 941 Dennison Building, 500 Church Street, Ann Arbor,
MI 48109, USA; \\
\supit{b} National Radio Astronomy Obs., 520 Edgemont Road, Charlottesville, VA
22903, USA;\\
\supit{c} Jet Propulsion Lab., California Institute of Technology, Pasadena, CA
91109, USA;\\
\supit{d} National Optical Astronomy Observatory, Tucson, AZ 85726-6732, USA;\\
\supit{e} Max-Planck Institute for Radio Astronomy, 69 Auf dem H\"ugel, Bonn,
Germany;\\
\supit{f} CRAL, Observatoire de Lyon, 9 av. Charles Andre, F-69561 Saint
Genis Laval Cedex, France;\\
\supit{g} Laboratoire Lagrange, Univ. de Nice, CNRS, Observatoire de la C\^ote
d'Azur, Nice, France;\\
\supit{h} University of Cambridge, JJ Thompson Avenue, Cambridge, CB3 0HE UK;\\
\supit{i} National Radio Astronomy Obs., Array Operation Center,
Socorro, NM 87801-0387, USA; \\
\supit{j} European Southern Obs. Casilla 19001, Santiago 19, Chile.
}

\authorinfo{Further author information: (Send correspondence to F. B.) \\
F.B.: E-mail: fbaron@umich.edu, Telephone: 1 734 615 4714}

\begin{document} 
 \maketitle 

\begin{abstract}
We present the results of the fifth Interferometric Imaging Beauty
Contest. The contest consists in blind imaging of test data sets derived from
model sources and distributed in the OIFITS format. Two scenarios of imaging
with CHARA/MIRC-6T were offered for reconstruction: imaging a T~Tauri disc and
imaging a spotted red supergiant. There were eight different
teams competing this time: Monnier with the software package MACIM;  Hofmann,
Schertl and Weigelt with IRS; Thi\'ebaut and Soulez with MiRA ; Young with
BSMEM; Mary and Vannier with MIROIRS; Millour and Vannier with independent BSMEM
and MiRA entries; Rengaswamy with an original method; and
Elias with the radio-astronomy package CASA.
The contest model images, the data delivered to the contestants and the
rules are described as well as the results of the image reconstruction obtained
by each method. These results are discussed as well as the strengths and
limitations of each algorithm.
\end{abstract}


\section{INTRODUCTION}
\label{sec:intro} 

The IAU Interferometry Beauty Contest is a competition aimed at encouraging
the development of new algorithms in the field of interferometric
imaging, by showcasing the current performance of image reconstruction
packages. The contest is being conducted by the Working Group on Image
Reconstruction of IAU Commission 54.

The principle of the contest is the following. One or several science cases
are first selected by the organizers, then realistic models of the science
targets are used to generate synthetic images. These ``truth'' images are then
turned into data sets, by simulating the acquisition of interferometric
observables by a typical interferometer. Finally, the contestants attempt to
reconstruct images from the data sets without knowledge of the original truth
images beyond the nature of the target. The reconstruction closest to the truth
image is then declared the winner.

The previous contests took place in 2004\cite{BC2004}, 2006\cite{BC2006},
2008\cite{BC2008} and 2010\cite{BC2010}, thus the 2012 Interferometry Beauty
Contest described here is the fifth contest. The contest results were
announced on July 5th during the 2012 SPIE Astronomical Telescopes and
Instrumentation conference in Amsterdam. 

\section{CONTEST MODEL, DATA AND GUIDELINES}

\subsection{Original model images}

In this 2012 edition of the Interferometry Beauty Contest, the focus was to
assess the potential of current reconstruction packages on very resolved
objects, under realistic observing scenarios. The organizers identified two
science cases for which the reconstruction performance is deemed critical for
interpretation: imaging Young Stellar Objects and imaging spotted stars.
Consequently two models were
generated for the Contest. A T~Tauri ``star + disc'' system that was nicknamed
Alp~Fak, and a red supergiant with bright spots named Bet~Fak. 

\begin{figure}
\centering
\includegraphics[width=0.5\linewidth]{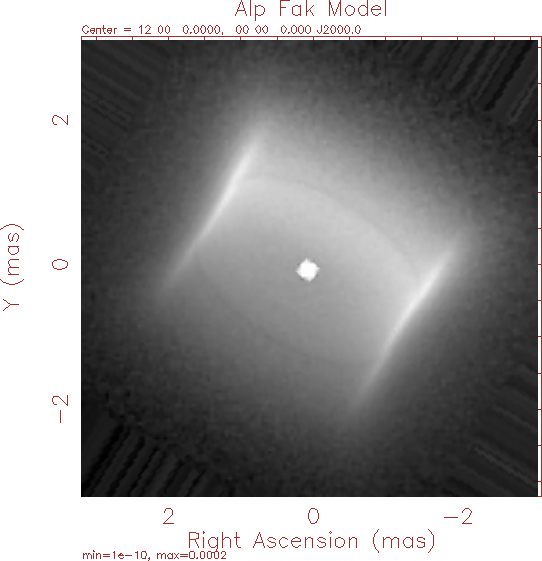}%
\includegraphics[width=0.5\linewidth]{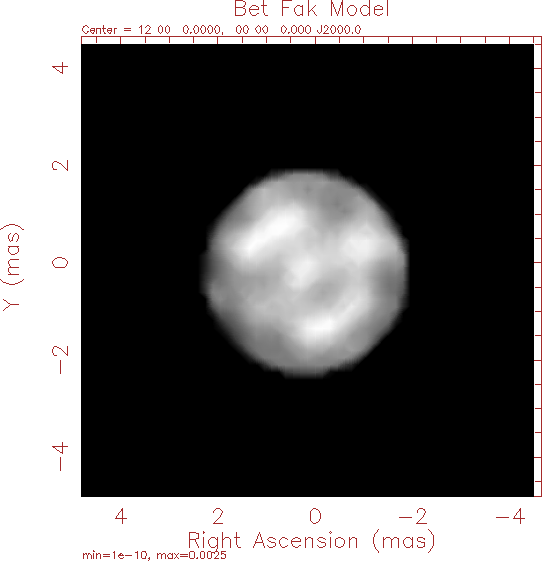}
\caption{Truth images: Alp Fak (left) and Bet Fak (right).}\label{fig:truth}
\end{figure}

The Alp~Fak image was modeled by Alicia Aarnio at the University of
Michigan, using the TORUS\cite{Harries2011} 3D radiative transfer code written
by Tim Harries at the University of Exeter.  The parameters for the T~Tauri
simulation were loosely based on that of v1295~Aql. The scaling of the image was
set to be slightly larger than the extent of the disc. The outer radius of the
disc was about $200$~AU, with the inner radius at $40$~AU. The central star, of
radius $3 R_{\odot}$, was offset by about $1.3$~AU in the $(X,Y)$ coordinates of
the image, but left in the midplane of the disk ($Z=0$). The mass accretion rate
was chosen low, as was the magnetospheric temperature, so that their effects on
the image are negligible. The resulting image was then rotated by $63.5^{\circ}$
to obtain the final truth image shown on Figure~\ref{fig:truth} (left). Our main
expectations from reconstructions of this object were: the detection of
the central source, a correct global orientation of the target, as well as
smooth flux and sharp transitions at the right locations. 

The Bet~Fak image is (to our knowledge) the first image in the Beauty Contest
that was partially derived from real data. This original data came from 2011
observations of the Red Supergiant AZ~Cyg, that presents clear asymmetric
features that may be spots or convection cells. Several images were
reconstructed from the original data, using complex combinations of regularizers
(total variation, $\ell_1\ell_2$ regularization and
limb-darkened disk prior of $3.9$~mas), and keeping the reduced $\chi^2$
below~$1.0$. A particularly ``good-looking'' image was picked amongst this one
to be the truth image of Bet~Fak, shown on Figure~\ref{fig:truth} (right).
Our main expectations from reconstructions of this object were: a smooth
circumference without artefacts (knowing that contestants would most likely use
priors), and approximate locations for the bright spots/convection cells.

\subsection{Data set generation: $(u,v)$ coverage and signal-to-noise}

Both data sets were simulated as if they were acquired by the
MIRC-6T\cite{Monnier2010} instrument (see Che et al., 2012 in these
proceedings), installed on the CHARA Array\cite{Brummelaar2005} atop Mt. Wilson,
CA, USA. The H-band low spectral
resolution mode of MIRC-6T was chosen for the simulations. In this mode the star
light is spectrally dispersed onto eight spectral channels. Because
very few software packages are able to handle multi-spectral reconstructions,
we chose not introduce any spectral dependency in the data. For a similar
reason, no temporal dependencies were assumed in the data beyond
aperture-synthesis
due to Earth's rotation. 

\begin{figure}
\centering
\includegraphics[width=0.45\linewidth]{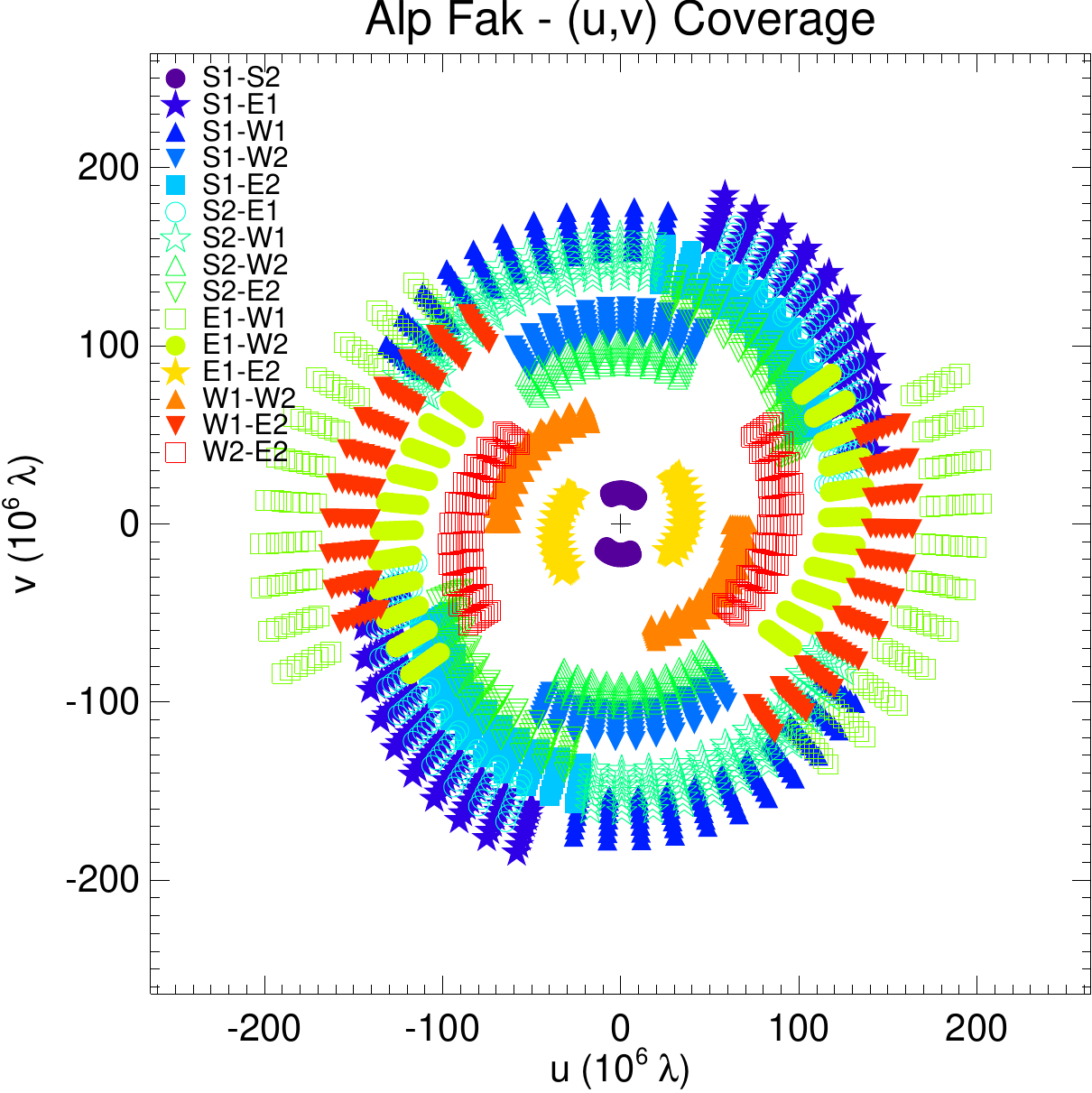}%
\includegraphics[width=0.45\linewidth]{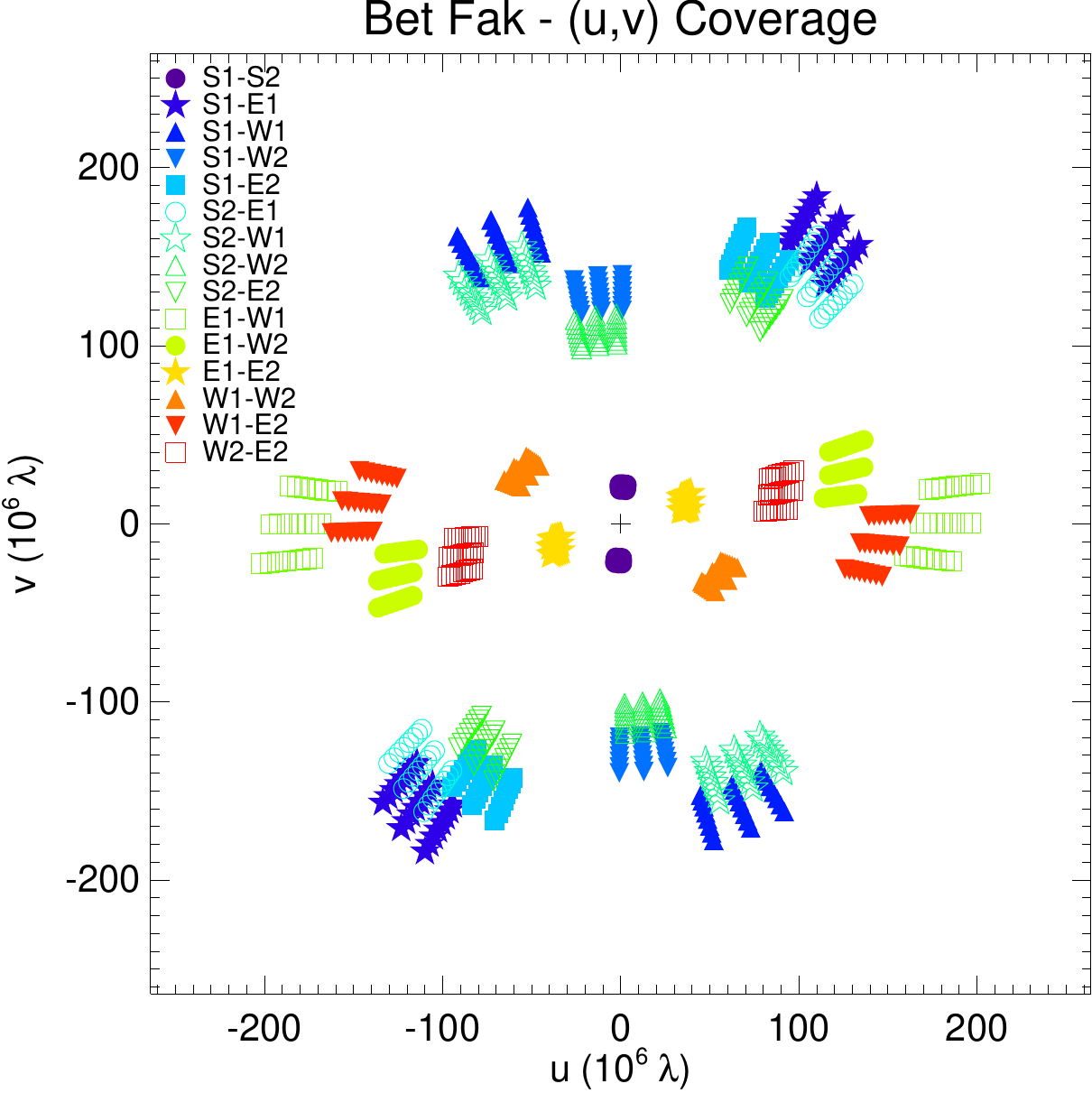}
\includegraphics[width=0.5\linewidth]{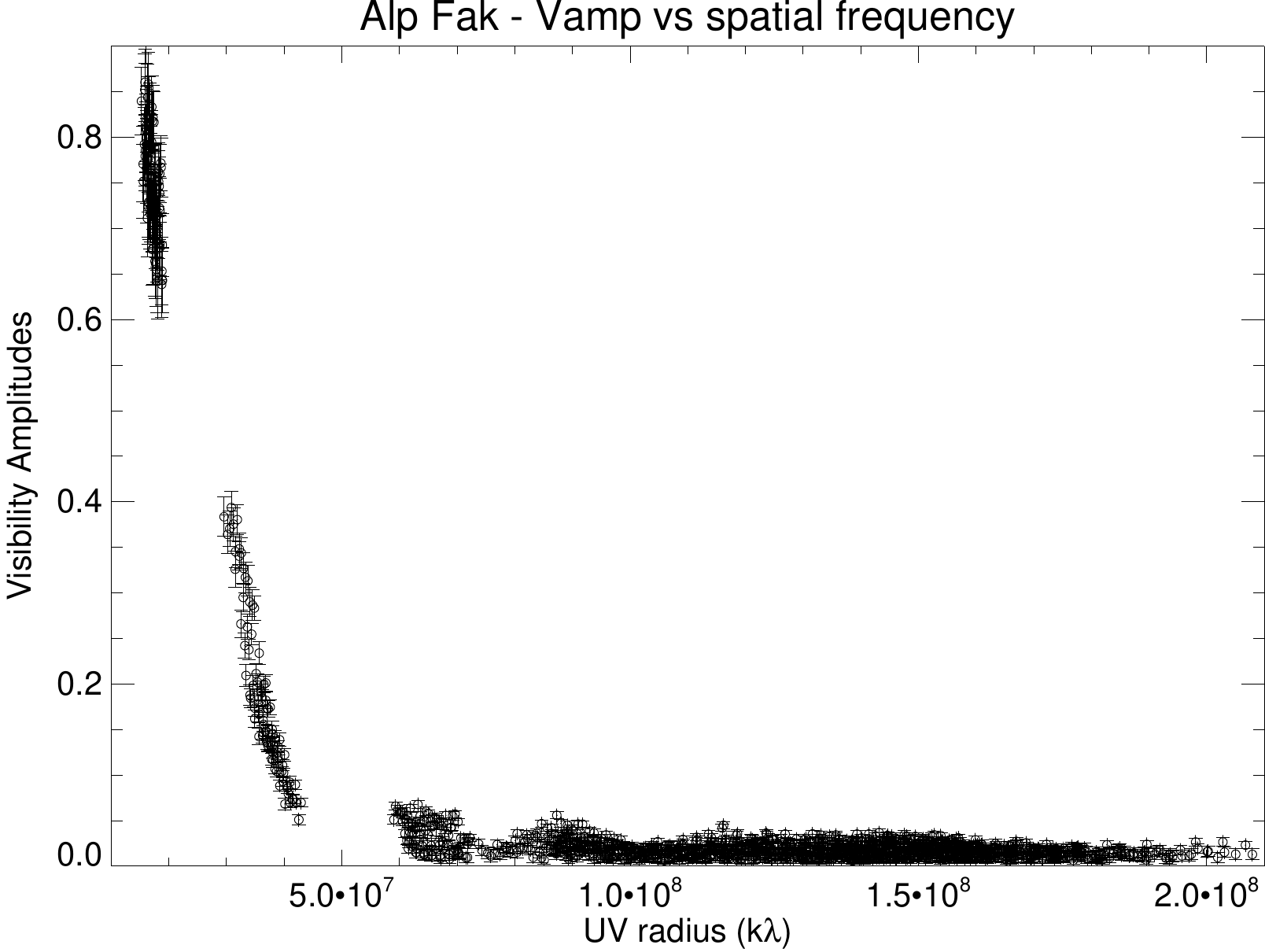}%
\includegraphics[width=0.5\linewidth]{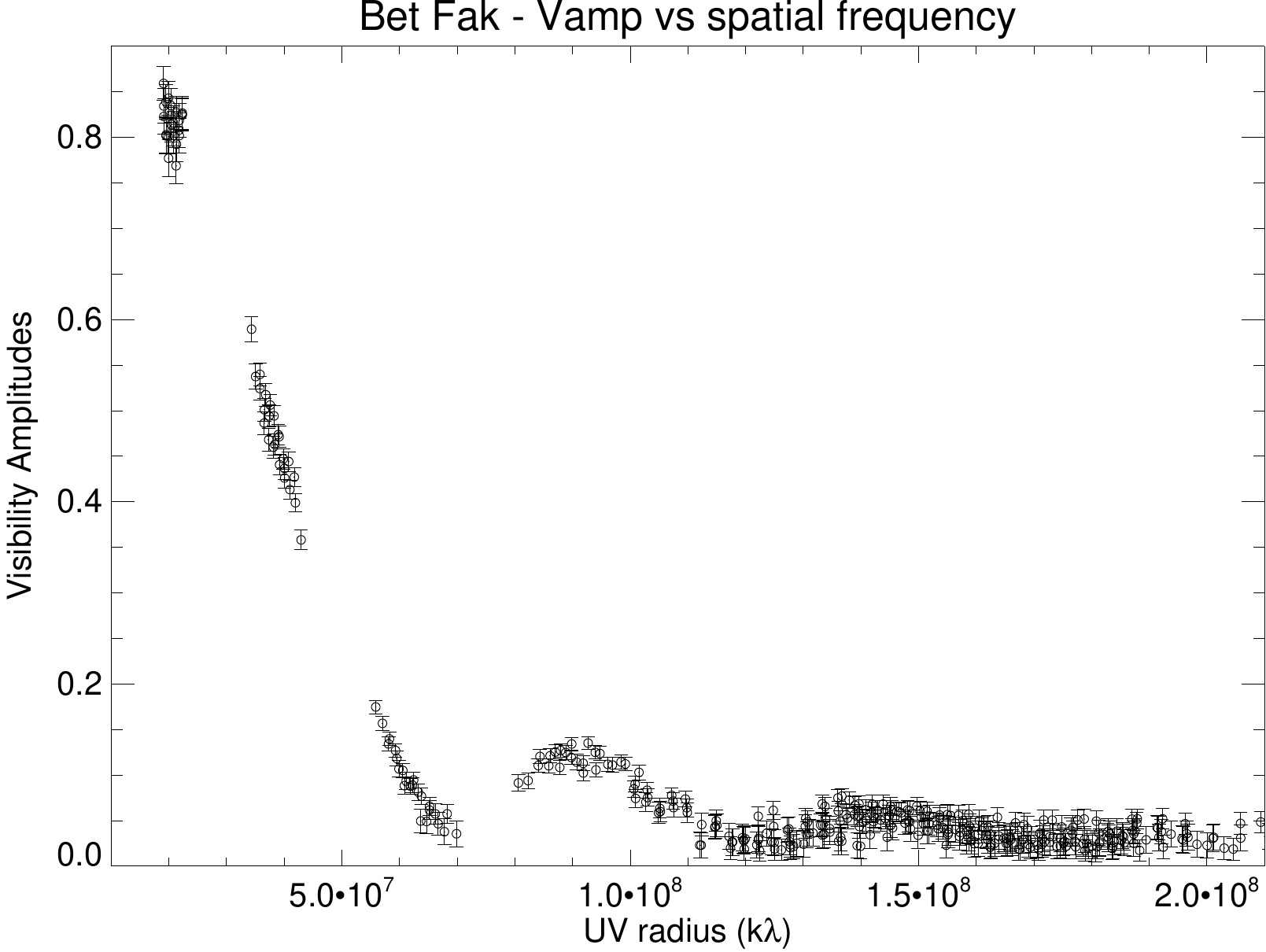}
\includegraphics[width=0.5\linewidth]{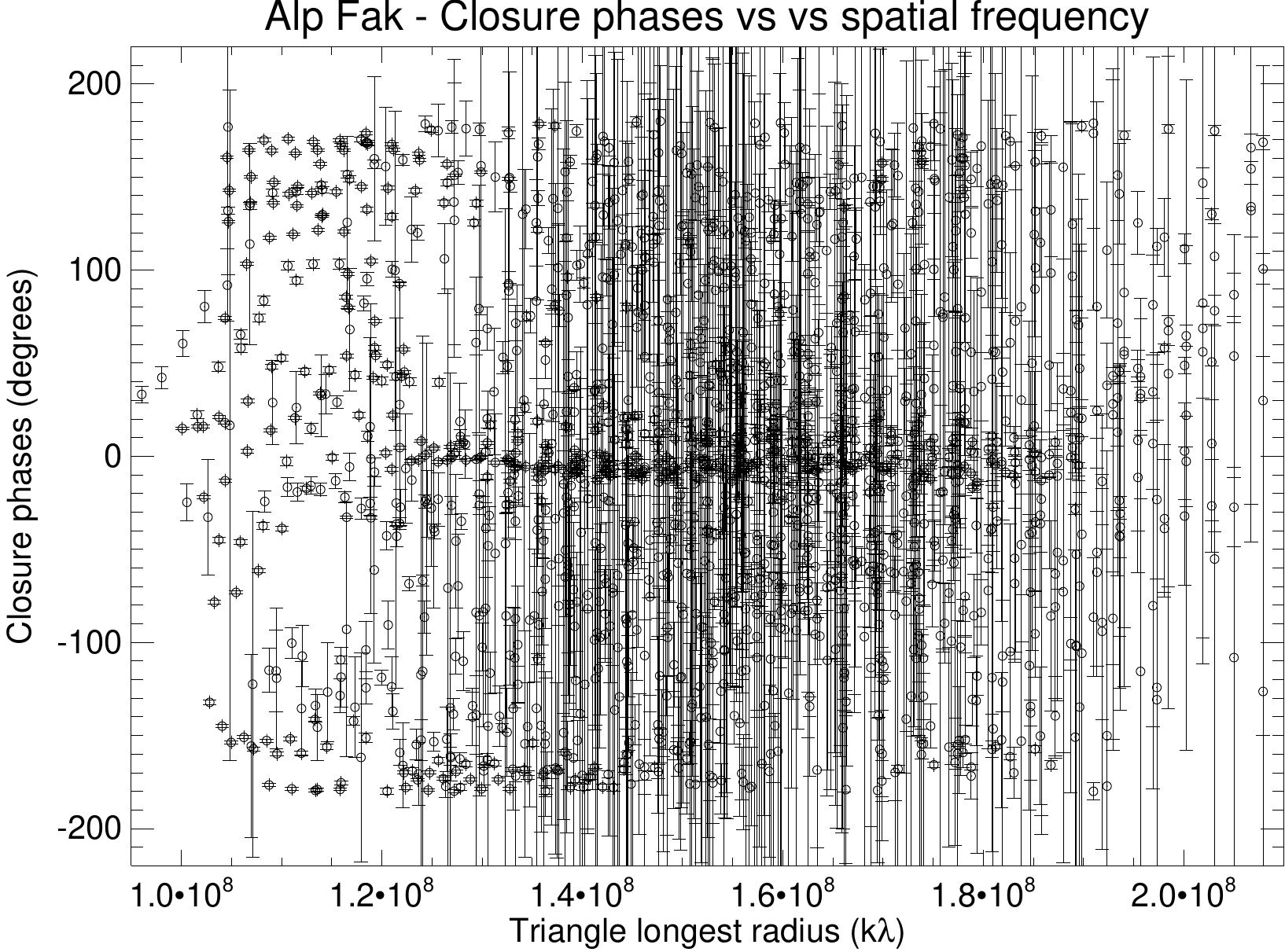}%
\includegraphics[width=0.5\linewidth]{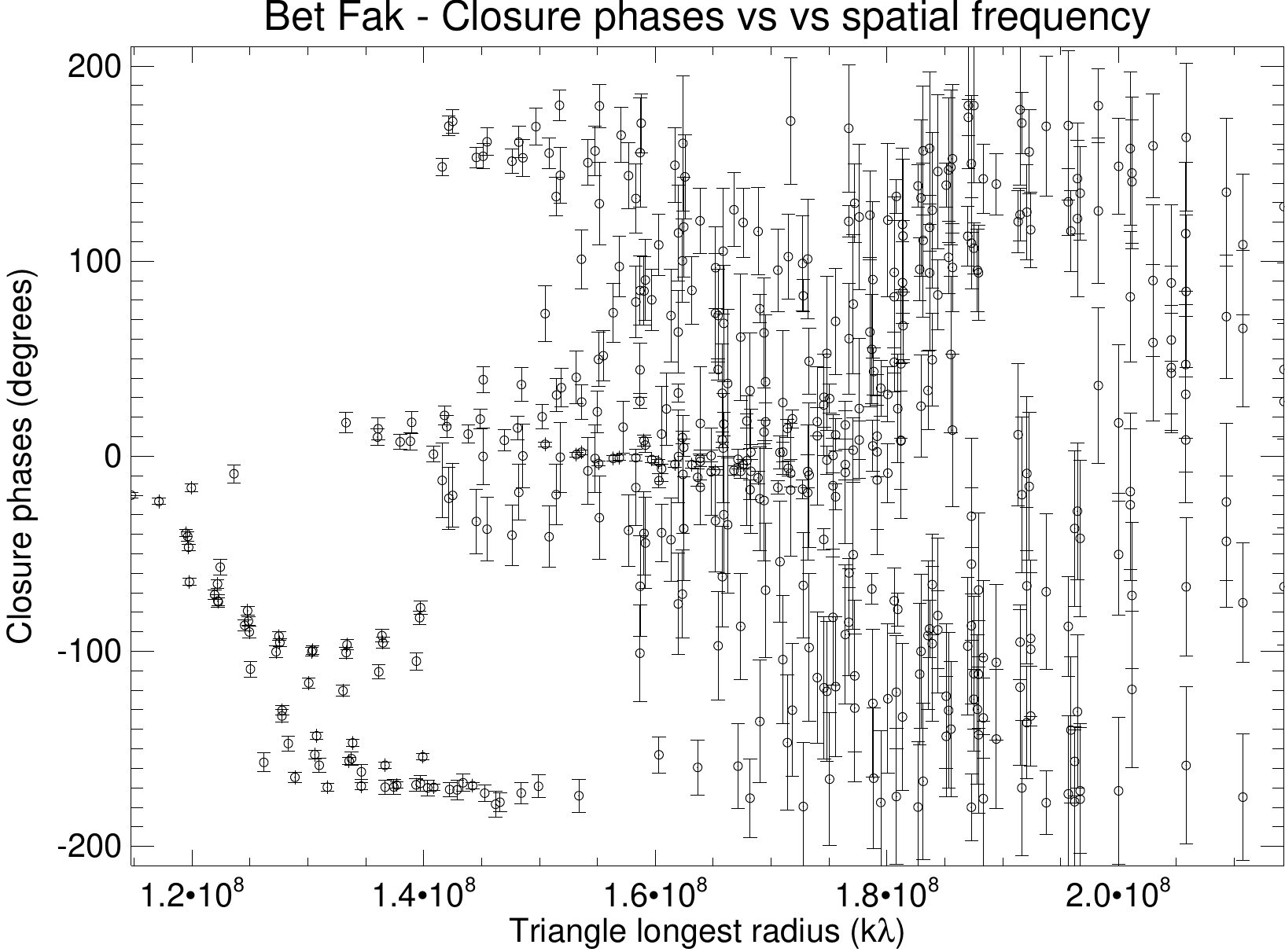}
\caption{Contest data characteristics for both targets Alp~Fak and Bet~Fak:
$(u,v)$ coverage (top), visibility amplitudes as a function of spatial
frequency (middle), and closure phases as a function of spatial frequency
(bottom). Note: because most squared visibilities are below $0.01$, visibility
amplitudes are shown here instead.} \label{fig:data}
\end{figure}

The $(u,v)$ coverage of the contest data is presented on
Figure~\ref{fig:data}. With MIRC-6T, the instantaneous ``snapshot'' Fourier
coverage provided is $15$ baselines -- i.e. $30$~$(u,v)$ points -- as well as
$10$ closure phases. For Alp~Fak, the complete $(u,v)$ coverage was chosen to
correspond to $5$~hours of observation, with snapshots acquired every
$15$~minutes. For Bet~Fak, the $(u,v)$ coverage was directly copied from the
original AZ~Cyg data. Compared to the previous Contest in 2010 that assumed the
use of $10$~VLTI stations, the $(u,v)$ coverage is much sparser.  In
particular, there is hole at low frequencies due to the
absence of CHARA baselines below $30$~m. On the other hand, the use
of six CHARA stations simulataneously (instead of e.g. only three with
VLTI/AMBER) allows to recover twice the amount of phase information.     
                                                                              
Once the Fourier sampling determined, the complex visibilities were computed via
Discrete Fourier Transform from the model images. The contest data consisted of
the conventional interferometric observables, i.e. power spectra (squared
visibilities) and bispectra (triple amplitudes + closure phases). These were
computed directly from complex visibilities, and modified using realistic
noises. For Alp~Fak our current best empiric noise model for MIRC-6T was
applied: typically a few percent errors on power spectra, $1^\circ$ to $5^\circ$
on closure phases for short baselines, and $10$ to $80^\circ$ on longer ones.
For Bet~Fak the signal-to-noise was directly copied from the original AZ~Cyg
data, thus reflecting the actual MIRC-6T noise. As shown on
Figure~\ref{fig:data},
visibility amplitude values are very low for both objects. In the case of
Alp~Fak, out of $1320$ power spectra, only $162$ are greater than $0.01$.

The noisy data were then packaged into OIFITS\cite{Pauls2005} data files, and
these were validated with the JMMC online validation tool
(\url{http://www.jmmc.fr/oival}).

Note that in addition to Alp~Fak and Bet~Fak, synthetic data of a binary
star with very high signal-to-noise and excellent $(u,v)$ coverage was provided.
This data was solely meant for contestants to
to check whether their software were able to reconstruct a simple model, and to
help them determine the orientation
of their reconstructions with respect to the default contest convention (North
up, East left). The binary separation was $4.0$~mas, with a principal
axis of $40^\circ$ East of North (from the bright star to the faint one) and a
flux
ratio of $5.0$. The uniform disk sizes were $1.0$~mas for the primary and
$0.75$~mas for the secondary.

Note that all these data sets (Alp Fak, Bet Fak, and the test binary)
are available on the Interferometry Beauty
Contest website\footnote{\url{
http://olbin.jpl.nasa.gov/iau/2012/Contest12.html}}.

\subsection{Contest guidelines}

In the past Beauty Contests, the contestants were free to choose the field of
view and pixel scale of their submissions. While this certainly made the
contests more challenging (in 2010, this freedom allowed the organizers to
``hide'' a point source far from the main target), this also raised the concern
that submissions had to be rescaled/convolved to the truth image pixellation. In
particular any super-resolution achieved by the reconstruction algorithms was
likely to be destroyed by such a procedure. Therefore, in this 2012 edition
of the Beauty Contest, both pixel scale and image
sizes were explicitly recommended. For both targets, the suggested pixel scale
was $0.15$~mas, with an advised field of view of $64 \times 64$ pixels.

\section{CONTEST SUBMISSIONS}

\begin{figure}
\centering
\begin{tabular}{ccc}
\hspace{5mm} Rengaswamy (unamed method) & \hspace{5mm} Elias (CASA) &
\hspace{5mm} Millour \& Vannier (BSMEM) \\
\includegraphics[width=0.3\linewidth]{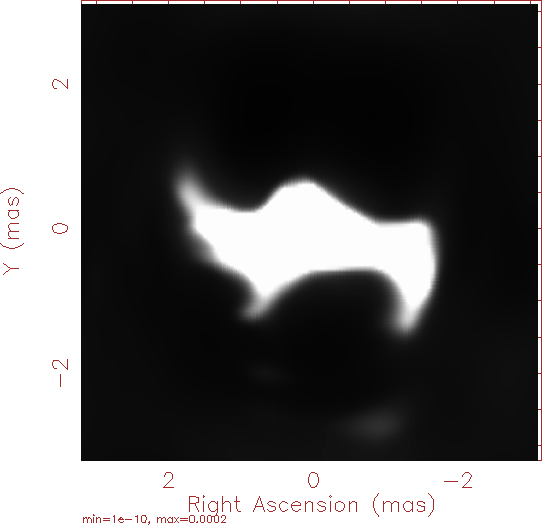}
& \includegraphics[width=0.3\linewidth]{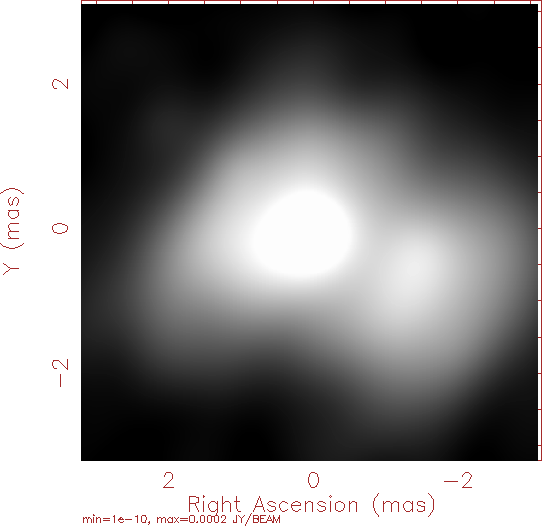}
&\includegraphics[width=0.3\linewidth]{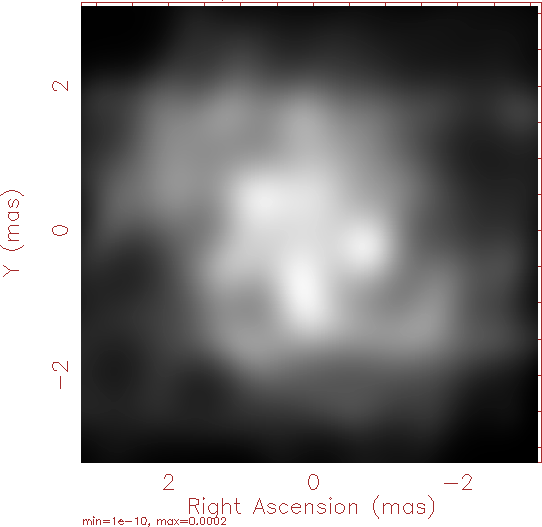} \\
\hspace{5mm} Young (BSMEM) & \hspace{5mm} Thi\'ebaut \& Soulez (MiRA) &
\hspace{5mm} Monnier (MACIM) \\
\includegraphics[width=0.3\linewidth]{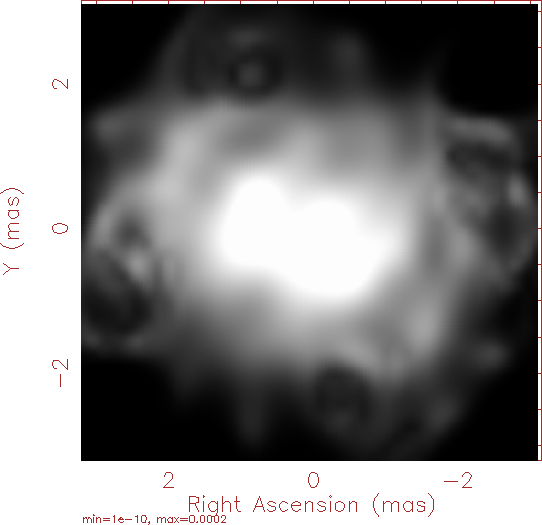}
&\includegraphics[width=0.3\linewidth]{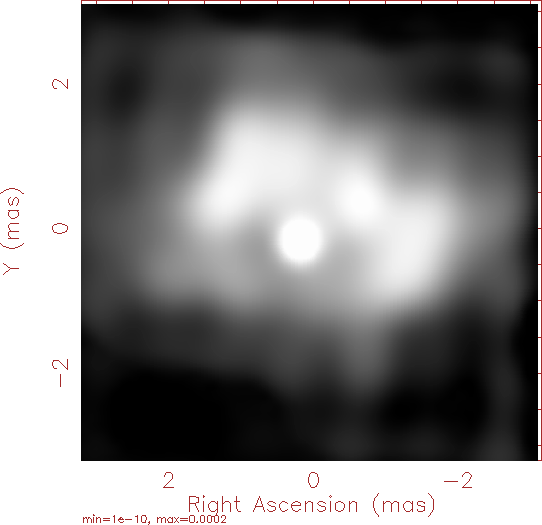}
&\includegraphics[width=0.3\linewidth]{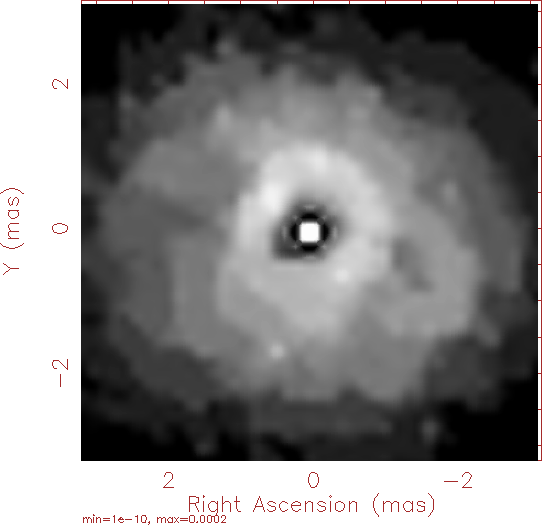} \\
\hspace{5mm} Mary \& Vannier (MIROIRS) &  \hspace{5mm} Millour \& Vannier (MiRA)
& \hspace{2mm} Hofmann, Schertl \& Weigelt (IRS) \\
\includegraphics[width=0.3\linewidth]{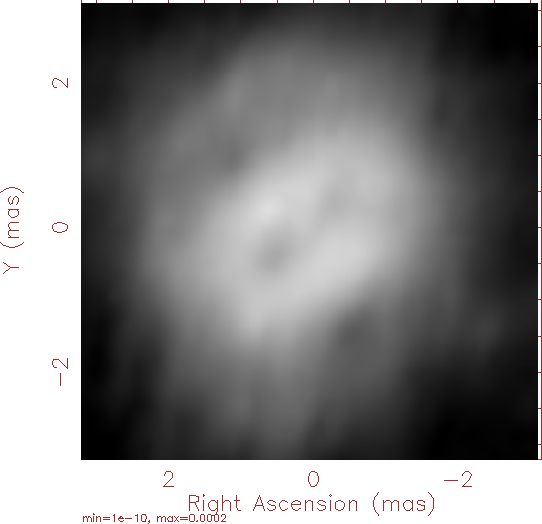}
&\includegraphics[width=0.3\linewidth]{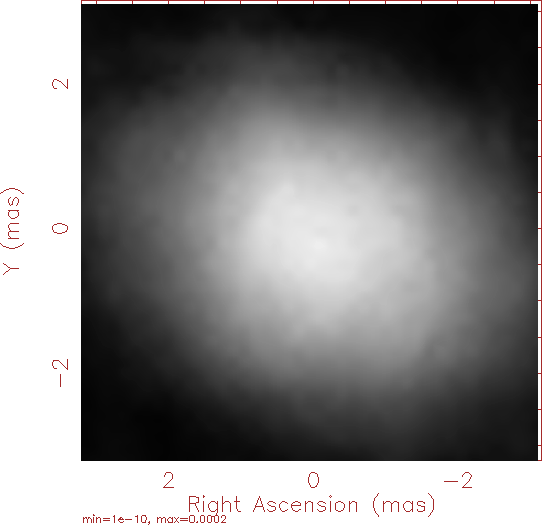}
&\includegraphics[width=0.3\linewidth]{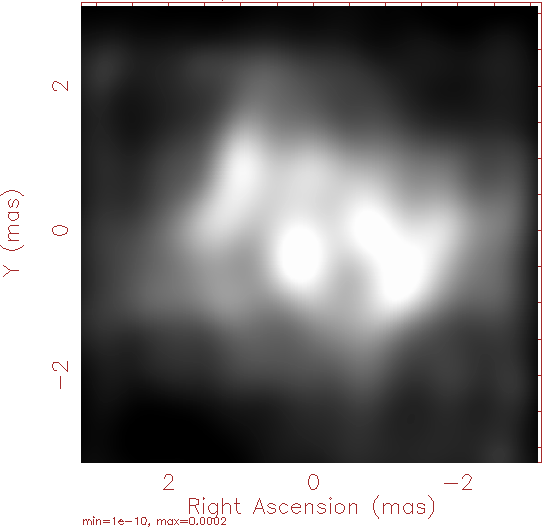} \\
\end{tabular}
\caption{Co-aligned contest submissions for Alp Fak, using the same greyscale
table as the truth image of Alp Fak presented on
Figure~\ref{fig:truth}.}\label{fig:alp_reconst}
\end{figure}

\begin{figure}
\centering
\begin{tabular}{ccc}
\hspace{5mm} Rengaswamy (unamed method) & \hspace{5mm} Elias (CASA) &
\hspace{5mm} Millour \& Vannier (BSMEM) \\
\includegraphics[width=0.3\linewidth]{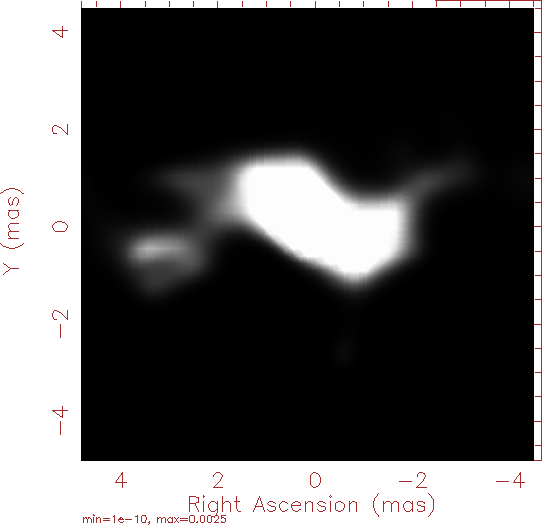}
& \includegraphics[width=0.3\linewidth]{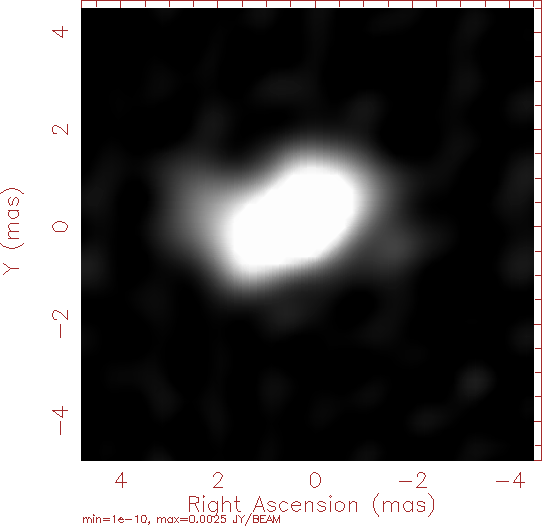}
&\includegraphics[width=0.3\linewidth]{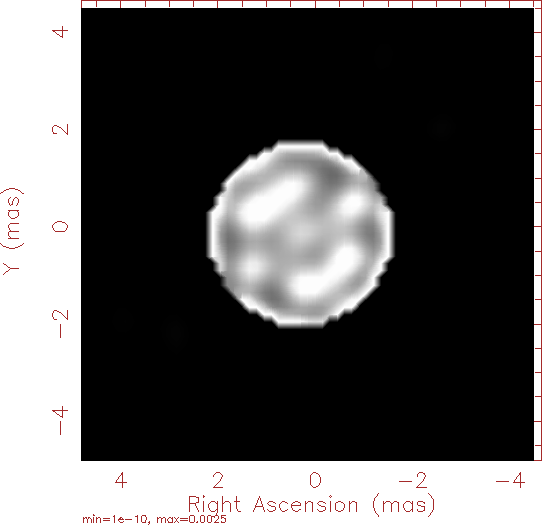}\\
\hspace{5mm} Young (BSMEM) & \hspace{5mm} Thi\'ebaut \& Soulez (MiRA) &
\hspace{5mm} Monnier (MACIM) \\
\includegraphics[width=0.3\linewidth]{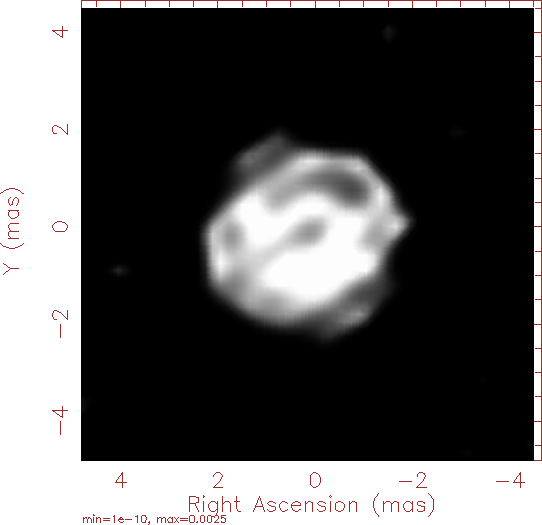}
&\includegraphics[width=0.3\linewidth]{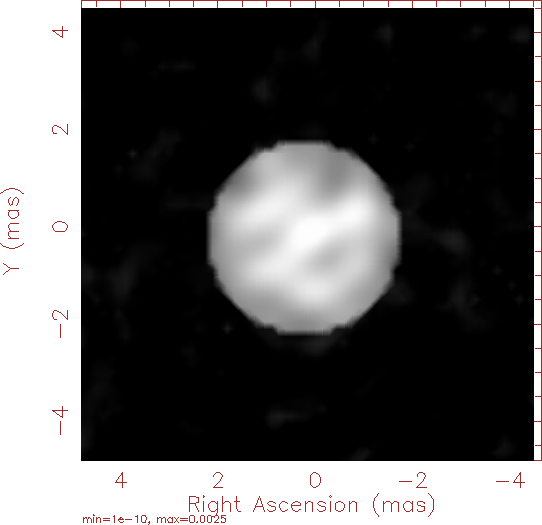}
&\includegraphics[width=0.3\linewidth]{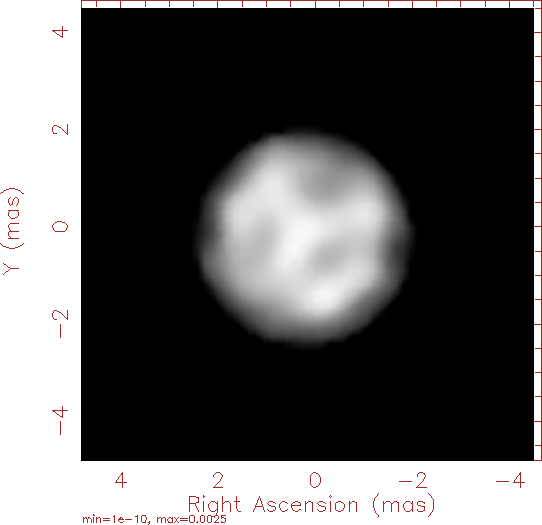}\\
\hspace{5mm} Mary \& Vannier (MIROIRS) &  \hspace{5mm} Millour \& Vannier (MiRA)
& \hspace{2mm} Hofmann, Schertl \& Weigelt (IRS) \\
\includegraphics[width=0.3\linewidth]{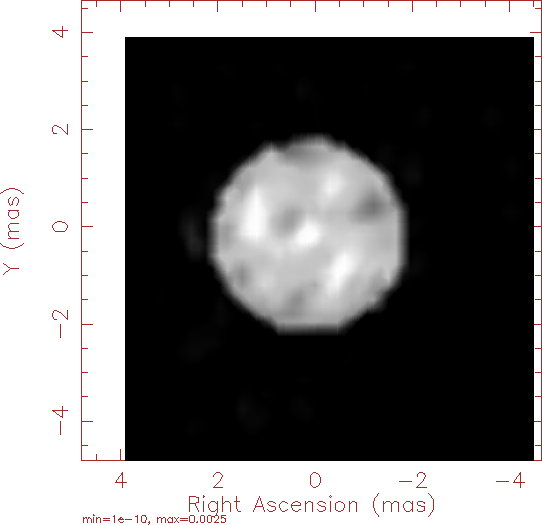}
&\includegraphics[width=0.3\linewidth]{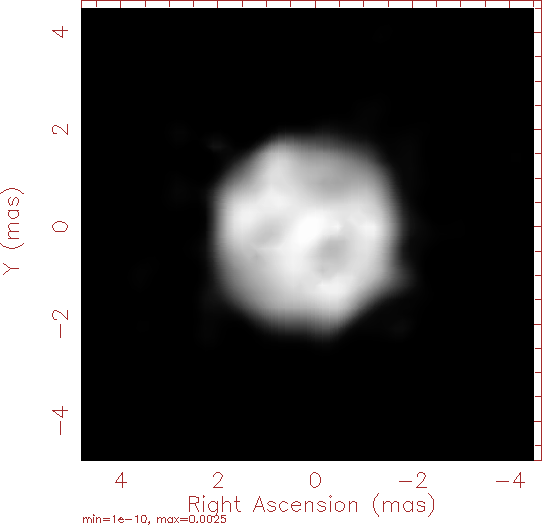}
&\includegraphics[width=0.3\linewidth]{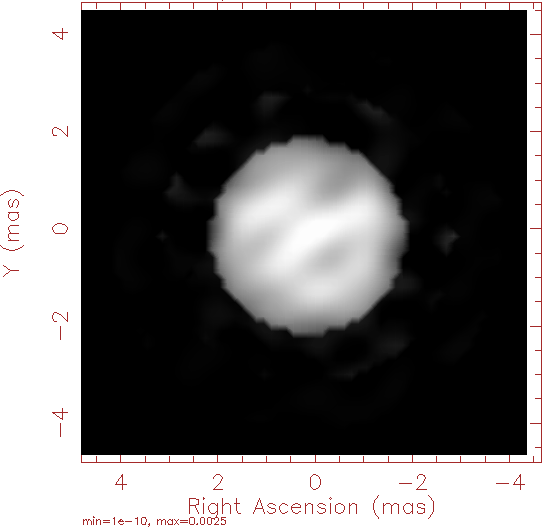}\\
\end{tabular}
\caption{Co-aligned contest submissions for Bet Fak, using the same greyscale
table as the truth image of Bet Fak presented on
Figure~\ref{fig:truth}.}\label{fig:bet_reconst}
\end{figure}

The 2012 Interferometric Imaging Beauty Contest enjoyed a record participation,
with eight teams in competition (compared to 4-5 for its previous editions).
In the following sections (3.1 to 3.8), the contestants present the
procedure they used for reconstruction, as well as the features they believe to
be real in their images. The reconstructions submitted by the contestants can be
found on Figure~\ref{fig:alp_reconst} for Alp~Fak and
Figure~\ref{fig:bet_reconst} for Bet~Fak. When more than one entry per team was
submitted for the same software, the best image was chosen for display.
Also, for the first time since the creation of the Beauty Contest, one team
(Millour \& Vannier) submitted  reconstructions with two packages they are not
actually developing (BSMEM and MiRA). Thus their approach was that of non-expert
but experienced users: it was particularly interesting in that it allows to
assess the similarities (or lack thereof) between reconstructions arising from
different user choices.

\subsection{MiRA}
\textbf{by Thi\'ebaut and Soulez (Observatoire de Lyon)}    

For both data sets Alpha Fak and Beta Fak, due to the lack of short
baseline on CHARA, the $(u,v)$ coverage is quite weak for low spatial
frequency. Consequently, the global shape of reconstructed objects is
quite difficult to recover, and strongly depends on the regularization and on
the starting solution (as the objective function is not convex due to the type
of interferometric data: squared visibilities and phase closures).

The first data set is a T Tauri star. As the squared visibilities as a function
of baseline doesn't show lobes, we supposed that the star is not resolved by the
interferometer. For that reason, we began the image restoration with a Dirac
smoothed to the resolution of CHARA (approx. $0.5$~mas = 3 pixels in the
restored image).  In accordance with Renard et al. \cite{Renard2011} (2011),
we choose total variation for the regularization. To avoid getting stuck in
local minima, we introduced some perturbations when the algorithm seemed to have
converged: e.g. by soft thresholding or by doing several
iterations using squared visibilities only, or squared visibilities and
closures only. We consider some kind of global convergence achieved when the
reduced $\chi^2$ is between $0.9$ and $1.0$ for the combination of square
visibilities and closures, square visibilities and triple amplitudes, and
square visibilities and bispectra. The reconstructed object presents a square
background of $6 \times 6$~mas that seems to be the size of the simulation box.
At the center of this squared background there is the unresolved bright star.
This star is surrounded by the half of an elliptic bright ring with a major axis
of about $3$~mas oriented at about $\text{PA} = 65^\circ$ (counted from North to
East).

The second data set, Beta Fak, is a red supergiant. As (u,v) coverage
is very sparse for low spatial frequencies, we estimated the global
shape of the star using a linear limb darkening model.  We fit the
parameters of this model by minimizing the same data cost function as
the one used by MiRA.  These parameters (diameter = 4.0 mas, limb
darkening parameter = 0.42) have been confirmed by LITPro\cite{TallonBosc2008}.
For the contest, we produced two images which have been
obtained from the square visibilities and closure phases (we did not use
triple amplitudes for this data set), starting with the limb darkening model and
with total variation or quadratic regularization.  This latter regularization is
taken as the
total quadratic difference between the image and the initial model.
In spite of using different regularizations, the two images are very
similar.  This is a (weak, because the problem is non-convex)
confirmation of the reality of the recovered structures at the stellar
surface.  The regularization weights have been tuned to have a final
normalized $\chi^2$ of $0.93$ per data sample.

\subsection{CASA}
\textbf{by Elias (NRAO)}

CASA\cite{Reid2010} is a radio interferometry package. It
requires visibility amplitudes and baseline phases, not the squared
visibilities and closure phases provided by the Beauty Contest organizers. CASA
can operate directly upon uvfits or measurement set (MS) format
files, not OIFITS format files. The MS is the native CASA file format.

Many radio interferometers employ ``fillers'' to convert their file
formats to MSes.  NME2 is in the process of creating an OIFITS to MS
filler within CASA now, but it is not yet ready.  For the beauty contest,
the OIFITS format files were converted to uvfits by the OYSTER
package\cite{Hummel2008} and in turn the uvfits files were converted to MSes by
CASA. OYSTER also estimated the baseline phases from closure phases before the
file format conversion.  For an array of six telescopes the number of
closures phases is $1/3$ less than the number of baseline phases, so the
system of equations is degenerate.  OYSTER determined the singular-value
decomposition (SVD) of the design matrix and formed the ``minimum-norm''
pseudo-inverse (infinite inverted singular values are set to zero).

The minimum-norm pseudoinverse is among the simplest and works well for
ensembles of a few point sources but not as well for extended emission. 
This technique will be available for the first version of the OIFITS to
MS filler. Additional model constraint inputs will eventually become
available. This task has the highest priority for the next Beauty Contest
as well as other advanced imaging techniques such as full-Stokes optical
interferometric polarimetry (OIP).

Once the Beauty Contest data were filled into CASA, they were imaged.  The
main imaging algorithm \cite{Rau2011} was an advanced version of CLEAN
employing multispatial scale (MSS) and multifrequency synthesis (MFS).  Standard
CLEAN determines and removes source components in the image plane using a
delta function, while MSS employs a configurable extended function. MFS
inserts data from all frequencies into a single uv plane before imaging. 
The frequency dependence for each pixel can be selected.  The Beauty
Contest data had flat spectra, so only a constant was fitted.

Several rounds of iterative CLEANing were interleaved with self-calibration,
where deviations between the model and observed visibilities
are considered to be calibration errors.  Many imaging trials were
performed with different CLEANing depths. Each trial led to a different
final image, which is symptomatic of ill-defined initial phases from the
minimum-norm pseudo-inverse.

\subsection{IRS}
\textbf{by Hofmann and Weigelt (Max Planck Institute f\"ur Radioastronomy)}

The iterative Image ReconStruction algorithm (IRS) uses the measured
bispectrum to reconstruct images. IRS uses the non-linear optimization
algorithm ASA-CG detailed in Hager \& Zhang \cite{Hager2005} (2005) and Hager
\& Zhang \cite{Hager2006} (2006). ASA-CG is a conjugate gradient based
algorithm. The advantage of IRS is that it is much faster than the Building
Block algorithm (Hofmann \& Weigelt \cite{Hofmann1993}, 1993).

The reconstructions are images of $64 \times 64$ pixels with a pixel scale of
$0.15$~mas and not convolved with a PSF. For the red supergiant Bet~Fak, we
reconstructed a disc with bright and dark spots. The intensities outside the
disk are probably artefacts. For the T Tauri disc Alp~Fak, we see an inclined
circumstellar disc with weak extensions above and below the inclined disc. The
long disc axis is approximately (but not exactly) horizontal.

\subsection{MIROIRS}
\textbf{by David Mary and Martin Vannier (University of Nice)}

We present here our contribution to the Beauty Contest, using the prototype
software MIROIRS (Methods for Image Reconstruction in Optical Interferometry
with Regularizations based on Sparse priors), which is still in a early phase.
Starting from scratch, our
progresses so far have led to a preliminary version based on the following
very simple principles. From an initial image (obtained using the LITpro
model-fitting software), we produce a gradient map of a criterion including the
target visibilities and phase closures. In this map, we consider a patch of
pixels (of possibly varying size, say, 5 by 5 to 20 by 20 pixels) around the
location of the strongest value of the gradient. This patch is used to define
the location of the image pixels where the flux should be changed to decrease
the criterion. The flux is changed by small increments which are proportional to
the gradient value within the patch, only for gradient values above a
predefined threshold. This gives the new image, from which the process is
iterated, with possible adjustments of the parameters (size of patch,
gradient threshold, increment) to ensure that the cost function is
iteratively decreased.

Our motivation for participating in the Beauty Contest was originally to
inject in the reconstruction algorithm models based on sparse
representations. This is done here in a very crude but fast way. We do not
impose any particular synthesis dictionary for the reconstruction. Instead,
the synthesis "atoms" are formed by analyzing and thresholding the gradient
of the cost function, and they come only as corrections to the initial
image, in a greedy manner. Optimization-wise, the adopted descent method is
very basic and largely empirical with respect to the tuning of the various
parameters involved. The relevance of the reconstructed image at the end of
the process also heavily depends on how close the initial image is to the
ideal solution. The prototype method is currently written in Matlab language. It
is not at all optimized and thus quite slow, but not prohibitively slow for the
present data set and ouput format.

As for our analysis of the presented result for the Beta Fak
reconstruction: we are confident that the circular structure is
reconstructed with a fairly correct diameter. The few tests we could make
starting from different initial images chosen as perturbed versions of a
uniform disk indicate that the reconstruction surface brightness
distribution is relatively stable under perturbations of the initial image.
This suggests that the bright and dark spots that are visible on the
surface may be real. We believe however that the faint structures around
the disk are reconstruction artefacts.

Concerning Alpha Fak, we also believe that the general shape we obtained is
not wrong (a large, elongated diffuse object oriented SW-NE), and  we think
that some internal structures might like look what we reconstructed. But,
here also, the $(u,v)$ coverage and the SNR at mid and high-frequencies does
not allow us to be very much confident about it.

\subsection{MACIM}
\textbf{by John Monnier (University of Michigan)}                               
                                                                              
I used MACIM using a ``uniform disk'' regularizer. which is the
$\ell_{\frac{1}{2}}$ norm of the spatial gradient of the image. Fabien Baron and
I invented this metric in 2011 -- it gives all uniform disks the same
regularization if they contain the same flux, irrespective of diameter. Just as
the total variation regularizer, this regularization prefers sharp boundaries,
but is agnostic on the size of the spot or feature. As for total variation, it
will prefer "round" spots compared to elliptical ones.
               
For the Alp Fak reconstruction. Based on inspection of the visibility curves and
expectation of a central unresolved source, I introduced a uniform disk model
containing 1.4\% of the flux. MACIM included this model component in the imaging
process, but the amount of flux in the point did not effect the regularization. 
This had the effect of creating a "hole" in the disk emission, likely coming
from a dust-free inner region in the model. I ran MACIM with a range of
regularizer weights and for different amounts of time, creating 3~different
possible images. I did not spend more than a few hours on this and chose the
image that showed approximately mean zero residuals for the short baselines,
which are easy to get
wrong because there are relatively few short baselines (i.e., one can iterate
too long and get a lower $\chi^2$ but the residuals show overfitting of long
baselines and underfitting of short baselines).  There are some intriguing
asymmetric structures in the disk but do not expect much of it is real. More
could
be done to symmetrize the images but I thought it was better to just leave the
MACIM result in a rather raw state.  For a real dataset, I would assess the
fidelity of features by using bootstrap methods or splitting up my data into
independent chunks.  Note that for my entry I just used a single pixel in the
center to represent the central star contribution, which is not exactly that
same model as MACIM used, but quite close. 

For the Bet~Fak reconstructions, I generate candidate limb-darkened disks (with
power law based on Lacour et al. 2008\cite{Lacour2008}, with coefficient $0.26$)
between $3.9$ and $4.30$~mas in diameter. I used these as weak priors in MACIM
along with ``uniform disc'' regularizer. I got similar structures in all
images, but found
the best agreement with the expected limb-darkening profiles and with the
short-baselines residual using the image reconstruction based on the $4.30$~mas
limb-darkening prior. I note that imaging spots of complex geometry is very
difficult. I would not publish this without additional independent datasets. I
also note that by increasing the regularizer strength I could smooth out most of
these structures with only a minor effect on the global $\chi^2$, emphasizing
the difficulty of this effort.

\subsection{Original (unnamed) method}
\textbf{by Sridharan Rengaswamy (European Southern Observatory)}    
 
The method used for reconstruction is the following:
\begin{itemize}
\item The closure phases (CP) are solved for visibility phases $\phi$ using
singular value decomposition (SVD) method. The phase solution is not unique. The
uncertainty in the solution is also estimated.

\item Complex visibilities are obtained as $\sqrt{V^2} \times \exp  i\phi$ and
weighted according to their signal-to-noise ratio.
\item The Dirty-map is obtained as the direct Fourier transform of the 2-d
visibility function. It is first multiplied by an apodization function in the 
image domain, Discrete Fourier transformed, multiplied by an apodization
function that is equivalent to the transfer function of a telescope
with diameter equal to the maximum baseline. The Dirty-beam is obtained in a
similar manner, replacing the complex visibilities at the measured uv points are
by unity.
\item The Dirty-map is then deconvolved with the dirty beam, using the Maximum
Entropy method (MEM). MEM iterations are stopped when the relative increase in
entropy is less than 1\% or when the entropy stats to decrease after reaching a
maximum.
\item Images were obtained in each spectral channel, as described in previous
steps. The final images were registered by cross-correlating the images with the
reference image (longest wavelength spectral channel image) and then added
together to obtain the final image.
\end{itemize}

The submitted images followed the Contest recommendations in terms of size and
pixellation. We found that only the morphology of the images (and not their
photometry) seem to be reliable. 

For Alp-Fak, the flux (sum of pixel
intensities) increases with the spectral channel (higher flux at longest
wavelength) alluding to the fact that the ‘object’ has a disk. Only
the central disk like structure (in log-scale) is reliable. The faint unresolved
features in the lower half at about $5.9$~and $8.7$~mas from the center are
unreliable. For Bet-Fak, the flux slowly increases but remains almost constant
at longer wavelengths (spectral channels). A bright granule of size $4.95 \times
2.5$ mas is clearly visible. There is also another small granule on its left,
separated by a dark lane. Other point like features in the images are not
reliable.

\subsection{BSMEM}
\textbf{by John Young (University of Cambridge)}    

The BSMEM (BiSpectrum Maximum Entropy Method) software was first
written in 1992 to demonstrate image reconstruction from optical
aperture synthesis data. It has been extensively enhanced and
tested since then, although there have been no changes of late. The
code used for this year's contest entry is essentially identical to
that employed for the 2010 contest. The algorithm applies a fully Bayesian
approach to the inverse problem of finding the most probable image given the
evidence, making use of the Maximum Entropy approach to maximize the posterior
probability of an image. An important advantage of BSMEM
is the automatic Bayesian estimation of the hyperparameter alpha that
controls the weighting of the entropic prior relative to the
likelihood. BSMEM can also perform a Bayesian estimation of missing
triple amplitudes and their associated errors from the power spectrum
data. BSMEM is available free-of-charge to the
scientific community on submission of the academic license agreement
at \url{http://www.mrao.cam.ac.uk/research/OAS/bsmem.html}.

BSMEM uses a trust region method with non-linear conjugate gradient
steps to minimize the sum of the log-likelihood ($\chi^2$) of the
data given the image and a regularization term expressed as the
Gull-Skilling entropy $\sum_k [I_k - M_k - I_k \log(I_k/M_k)]$. The
model image $M_k$ is usually chosen to be a Gaussian, a uniform disk, or
a delta-function centered in the field of view, which conveniently
fixes the location of the reconstructed object (the bispectra and
power spectra being invariant to translation).  This type of starting
model also acts as a support constraint by penalizing the presence of
flux far from the center of the image. 

The reconstruction of the T Tauri disk (Alp~Fak) used a circular
Gaussian default image (with FWHM found by fitting to the short-baseline
squared visibility data). For the reconstruction of the supergiant star (Bet
Fak) surface, I found it necessary to use additional prior information to
constrain the radial distribution of flux. This was obtained by fitting a
circular limb-darkened disk (Hestroffer model) to the squared
visibility data (elliptical models did not fit significantly
better). The image corresponding to the best-fit limb-darkened model
was convolved with a 0.3 mas FWHM Gaussian blur, in order to avoid
penalizing slight deviations of the disk edge from circular symmetry,
before being used as the default image in BSMEM. Following the advice of the
contest organizer, a pixel size of
0.15 mas was selected for both contest objects.

For the supergiant star (Bet Fak), I am confident the following
features are real: the three brightest spots (S, W, and E) in the central
region of the stellar surface; the protrusion at the NE edge of the star; the
"cut-out" at the W edge of the star. I am not convinced that the possible
companion with position angle ~$100^\circ$ E of N is real. Certainly all of
the other features outside of the star are artefacts.

For the T Tauri object (Alp Fak), I am confident in the overall shape
and orientation of the disk (quasi-rectangular outer contours and
elliptical inner contours). At the centre of the image there are
possibly two sections of a bright inner disk rim and a central star.

\subsection{BSMEM and MiRA}
\textbf{by Florentin Millour and Martin Vannier (University of Nice)}

We tried to place ourselves from a user point of view, i.e. we wished 
to use several (ideally most of) available image reconstruction 
software, and compare the obtained results. Therefore, our plans were 
initially to reconstruct images using image reconstruction 
software that are available: MiRA and BSMEM. We also tried to use the WISARD
software, using complementary low-frequency data derived from model-fitting.
However we noticed some incoherency in the way WISARD reads this data. This
probably stems both from some inconsistency in our home-made OIFITS file, and
from a lack of robustness of the WISARD conversion routine from the OIFITS
format. As this problem could not be tackled in time, we could only
obtained flawed reconstructions from WISARD, which we chose not to show here.

The presented MiRA and BSMEM images have 64 pixels, and 0.15 mas per pixel,
supposedly increasing with increasing alpha and delta. They are centered around
the photocenter of the image. Our method consisted in the following:
\begin{itemize}
\item We first fitted the datasets in the first visibility lobe using the 
softwares fitOmatic (home-made) and LITpro (developed by JMMC).

\item A synthetic OIFITS file containing squared visibilities (and closure 
phases) was generated out of this fit result, for each object, with 
uniformly and randomly spread baselines in this same first visibility 
lobe. This synthetic dataset was added to the original dataset. The idea 
here was to strongly limit the field of view of the image reconstruction 
to the effective size of the object.

\item from this point on, different procedures were used depending on the
image reconstruction software:
\begin{itemize}
   \item for MiRA, $300$~images were generated with a random initial 
guess (either a uniformly random image, a random-width Gaussian, or a 
random-width uniform disk). These were sorted by increasing $\chi^2$,
and any image with a $\chi^2$ larger than 2 times the minimum one was
discarded. All other images were kept and averaged in the result image. 
This first image run was used as a prior for a second run of $300$~generated
images, the same way. The result was averaged and produced the  presented
result. We estimate that the centering process involved in the averaging
decreases the effective resolution of the image to $0.45$~mas. For more
information on our procedure, see Millour \& Vannier (2012) in the same
proceedings.

\item For BSMEM, we used the default method and reconstruction 
parameters. We just considered for Alp~Fak a Gaussian prior with the size 
of the fitted model we have done previously (3 mas), and for Bet~Fak a 
uniform disk of diameter 3.5 mas. There was no subsequent 
convolution with a beam, so the resolution is supposedly $0.15$~mas (whereas the
effective details would show up at a resolution of approximately~$1$ mas).
\end{itemize}
\end{itemize}

Our interpretation of these datasets is the following:
\begin{itemize}
\item for Alp~Fak: the model fitting provides an overall elongated Gaussian
shape 
with major axis 4.2 mas, minor axis 3.1 mas, and position angle 66 
degrees. The attempts to reconstruct images (and also the model fitting) 
indicate that there are other, finer structures, like 2 or 3 "clumps" 
near the Gaussian center, but we were unable to locate them clearly. Our 
guess would be one clump in the center (the star) plus 2 clumps on both 
sides, along the major axis, which would represent an hypothetical inner 
rim.
\item for Bet~Fak: The model fitting (and the shape of visibilities vs spatial 
frequencies) indicates us a circular uniform disk plus 1 or 2 bright 
spots. Indeed, we are able to get spots in the reconstructed images, but 
we believe only MiRA is able to locate them correctly, whereas BSMEM tends to
produce symmetric images of these spots.
\end{itemize}

\section{COMPARISION METRICS AND CONTEST RESULT}

\begin{table}[t]
\centering
\begin{tabular}{|c|c|c|c|c|}
\hline Team & Software & Alp Fak score & Bet Fak score & Total score \\ \hline
Monnier                    &  MACIM &  37.8 & 242.5  & 280 \\
Hofmann, Schertl \& Weigelt & IRS &  28.2 & 285.1  & 313\\ 
Millour \& Vannier         & MiRA & 27.7 & 299.8 & 327 \\
Thi\'ebaut \& Soulez         & MiRA  & 24.8 & 337.6 & 362 \\
Mary \& Vannier            & MIROIRS & 29.4 & 381.9 & 411 \\
Young                      & BSMEM   & 40.2 & 659.9 & 700 \\
Millour \& Vannier & BSMEM &  32.3 & 871.7 & 903 \\
Elias & CASA &  41.2 & 1285.9 & 1327\\
Sridharan Rengaswamy & SR & 235.6 & 1636.6  & 1872 \\ \hline
\end{tabular}
\caption{Official results of the Beauty Contest 2012.}\label{tab:results}
\end{table}

The judge of the 2012 Interferometric Imaging Beauty Contest was William Cotton.
The scoring of the Beauty Contest data was done through the following (and now
standard) procedure:
\begin{enumerate}
\item the submissions are aligned using features in the images. This is done by
correlating the submissions with the truth image. This also includes image
flips if required, as contestants did not necessarily use the Beauty Contest
orientation conventions. The Beauty Contest adopts the standard convention for
optical interferometry: East to the left, and North up. 
\item the submitted images are interpolated onto the pixellation grid of the
master images, if needed. For contestants that submitted images at the
recommended pixellation of $0.15$~mas, this step was skipped.
\item All images (including the truth images) are normalized to unity in a
given box. For Alp~Fak this was the rectangle defined by corners $[7,7]$ and
$[119,119]$, representing more than $90\%$ of the emission. For Bet~Fak this
was the box within $14$~pixels of the center, that included essentially all
emission.
\item For each object, the score is computed as $10^6$ times the RMS
pixel-by-pixel difference between the submission and the truth image in the
boxes defined in 3).
\item The scores for Alp~Fak and Bet~Fak are added. The best scores are the
lowest.
\end{enumerate}
If more than one image was submitted per object and per team, the one achieving
the best score was retained for scoring (e.g. Thi\'ebaut and Soulez
submitted images of Bet~Fake regularized with total variation and a quadratic
regularizer).

\begin{figure}[t]
\centering
\includegraphics[width=0.7\linewidth]{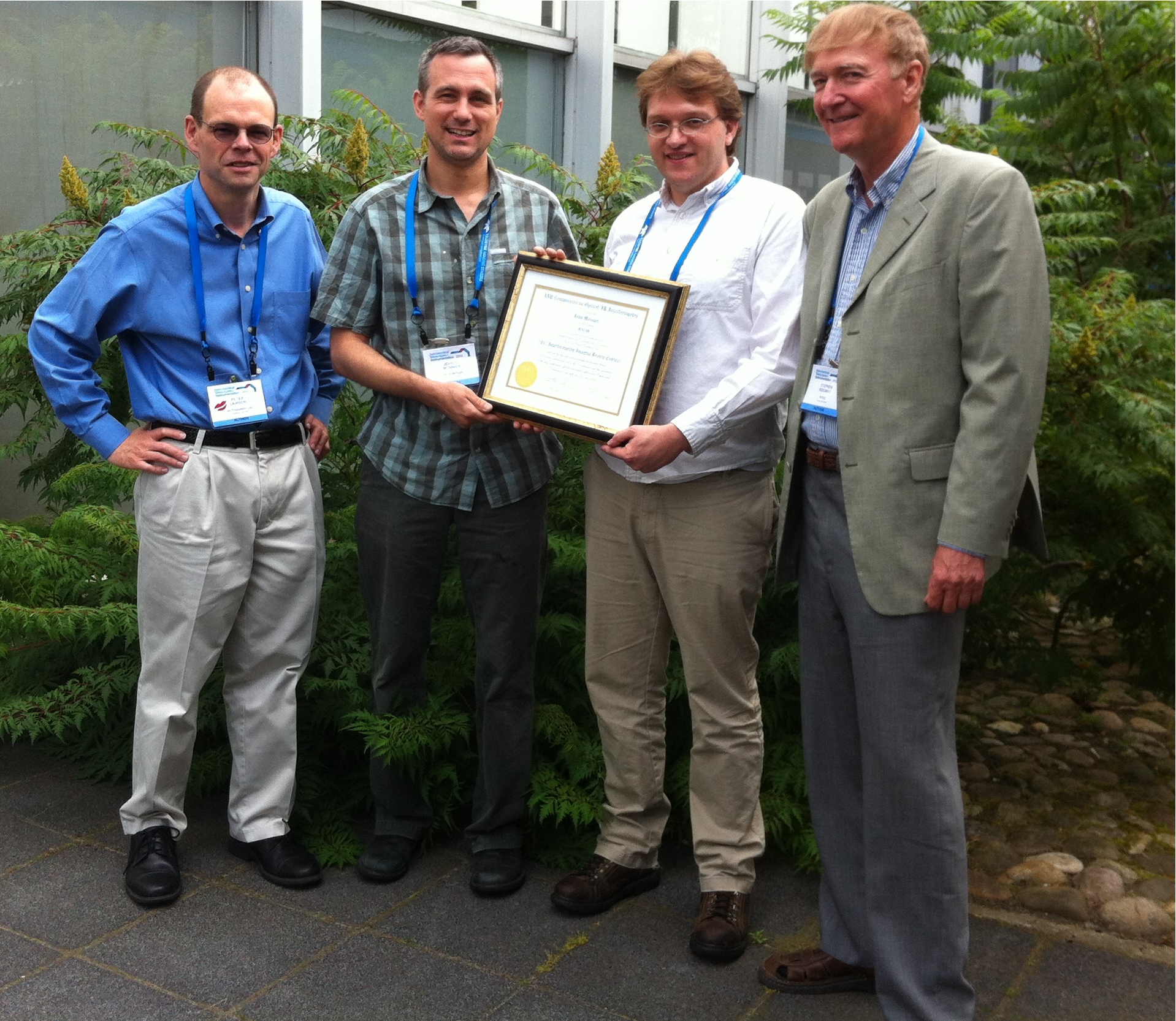}
\caption{ The 2012 Interferometry Imaging Beauty Contest jury presenting
the award to the winner. From left to
right: P. Lawson, J. Monnier (2012 contest winner), F.
Baron, S. Ridgway.}\label{fig:award}
\end{figure}

The official 2012 Interferometric Imaging Beauty Contest scores are given in
Table~\ref{tab:results}. In light of the results, John Monnier (University of
Michigan) was declared the winner of the contest for his MACIM entry and
was awarded the contest prize by the jury (see Figure~\ref{fig:award}).

\section{DISCUSSION AND CONCLUSION}

The general agreement amongst contestants is that both targets were hard
to reconstruct. This gave rise to more variance in reconstruction quality
than what was witnessed during previous contests; but also demonstrated
that decent imaging quality can be obtained on very resolved objects in
realistic conditions.

Alp~Fak was a difficult target, being probably too resolved for any
current software to reconstruct it very well. MiRA and IRS obtained the best
results, managing to reconstruct a smooth central star. The regularization used
by MACIM favored uniform patches of fluxes, and thus was most probably not
adapted to recover the original distribution.

Bet~Fak was overall reconstructed well in terms of size, but the actual location
of the spots was very dependent on the algorithm. Perhaps because stellar
surface imaging is a more familiar application, image priors such as
limb-darkened disks were used by most reconstruction. As both the $(u,v)$
coverage and signal-to-noise of Bet~Fak were derived from real CHARA/MIRC-6T
data, these results demonstrate caution will be needed when reconstructing spots
from real data. MiRA achieved the best scores on
Alp~Fak, but its lower performance on Bet~Fak (as well as the current
specificities of the Beauty Contest metrics that put more weight on this target)
prevented it from getting the best overall scores. 

With record participation and overall convincing reconstructions,
most contestants
felt that the fifth Beauty Contest was successful at showcasing the
diversity and strengths of the current imaging packages in monochromatic mode.
As several packages are planed to add multi-wavelength imaging capabilities in
2012-2013, this contest may be indeed the last one to figure only
monochromatic data. As multi-wavelength image reconstruction is both
a difficult algorithmic problem and a necessity for new science, the next Beauty
Contests should definitively prove exciting... 

\acknowledgments  
Work by Fabien Baron was supported by the National Science Foundation through
awards AST-0807577 to the University of Michigan. Work by Peter R. Lawson was
undertaken at the Jet Propulsion Laboratory, California Institute of Technology,
under contract with the National Aeronautics and Space Administration. Work by
William D. Cotton was supported by the National Radio Astronomy Observatory, a
facility of the National Science Foundation operated under cooperative agreement
by Associated Universities, Inc..


\bibliography{bc2012} 

\begin{thebibliography}{10}

\bibitem{BC2004}
{Lawson}, P.~R., {Cotton}, W.~D., {Hummel}, C.~A., {Monnier}, J.~D., {Zhao},
  M., {Young}, J.~S., {Thorsteinsson}, H., {Meimon}, S.~C., {Mugnier}, L., {Le
  Besnerais}, G., {Thiebaut}, E., and {Tuthill}, P.~G., ``{The 2004 Optical/IR
  Interferometry Imaging Beauty Contest},'' {\em \procspie} {\bf 5491},  886
  (2004).

\bibitem{BC2006}
{Lawson}, P.~R., {Cotton}, W.~D., {Hummel}, C.~A., {Baron}, F., {Young}, J.~S.,
  {Kraus}, S., {Hofmann}, K.-H., {Weigelt}, G.~P., {Ireland}, M., {Monnier},
  J.~D., {Thi{\'e}baut}, E., {Rengaswamy}, S., and {Chesneau}, O., ``{2006
  interferometry imaging beauty contest},'' {\em \procspie} {\bf 6268} (2006).

\bibitem{BC2008}
{Cotton}, W., {Monnier}, J., {Baron}, F., {Hofmann}, K.-H., {Kraus}, S.,
  {Weigelt}, G., {Rengaswamy}, S., {Thi{\'e}baut}, E., {Lawson}, P., {Jaffe},
  W., {Hummel}, C., {Pauls}, T., {Schmitt}, H., {Tuthill}, P., and {Young}, J.,
  ``{2008 imaging beauty contest},'' {\em \procspie} {\bf 7013} (2008).

\bibitem{BC2010}
Malbet, F., Cotton, W., Duvert, G., and Lawson, P., ``{The 2010 interferometric
  imaging beauty contest},'' {\em \procspie}~{\bf 7734},  77342 (2010).

\bibitem{Harries2011}
{Harries}, T.~J., ``{An algorithm for Monte Carlo time-dependent radiation
  transfer},'' {\em MNRAS}~{\bf 416},  1500--1508 (Sept. 2011).

\bibitem{Monnier2010}
{Monnier}, J.~D., {Anderson}, M., {Baron}, F., {Berger}, D.~H., {Che}, X.,
  {Eckhause}, T., {Kraus}, S., {Pedretti}, E., {Thureau}, N., {Millan-Gabet},
  R., {Ten Brummelaar}, T., {Irwin}, P., and {Zhao}, M., ``{MI-6: Michigan
  interferometry with six telescopes},'' {\em \it Society of Photo-Optical
  Instrumentation Engineers (SPIE) Conference Series} {\bf 7734} (July 2010).

\bibitem{Brummelaar2005}
ten Brummelaar, T., McAlister, H., Ridgway, S., Bagnuolo, W.~J., Turner, N.~H.,
  Sturmann, L., Sturmann, J., Berger, D.~H., Ogden, C.~E., Cadman, R.,
  Hartkopf, W.~I., Hopper, C.~H., and Shure, M.~A., ``{First results from the
  CHARA Array. II. A description of the instrument},'' {\em The Astrophysical
  Journal}~{\bf 628},  453 (July 2005).

\bibitem{Pauls2005}
{Pauls}, T.~A., {Young}, J.~S., {Cotton}, W.~D., and {Monnier}, J.~D., ``{A
  Data Exchange Standard for Optical (Visible/IR) Interferometry},'' {\em The
  Publications of the Astronomical Society of the Pacific}~{\bf 117},
  1255--1262 (Nov. 2005).

\bibitem{Renard2011}
Renard, S., Thi\'{e}baut, E., and Malbet, F., ``{Image reconstruction in
  optical interferometry: benchmarking the regularization},'' {\em \aap}~{\bf
  533},  A64 (Aug. 2011).

\bibitem{TallonBosc2008}
{Tallon-Bosc}, I., {Tallon}, M., {Thi{\'e}baut}, E., {B{\'e}chet}, C., {Mella},
  G., {Lafrasse}, S., {Chesneau}, O., {Domiciano de Souza}, A., {Duvert}, G.,
  {Mourard}, D., {Petrov}, R., and {Vannier}, M., ``{LITpro: a model fitting
  software for optical interferometry},'' {\em \it Society of Photo-Optical
  Instrumentation Engineers (SPIE) Conference Series} {\bf 7013} (July 2008).

\bibitem{Reid2010}
{Reid}, R.~I. and {CASA Team}, ``{CASA: Common Astronomy Software
  Applications},'' in [{\em American Astronomical Society Meeting Abstracts
  215}{\nolinebreak\hspace{0.1em}]},  {\em Bulletin of the American
  Astronomical Society} {\bf 42},  568 (Jan. 2010).

\bibitem{Hummel2008}
{Hummel}, C.~A., ``{QC and Analysis of MIDI Data Using mymidigui and OYSTER},''
  in [{\em 2007 ESO Instrument Calibration
  Workshop}{\nolinebreak\hspace{0.1em}]},  {Kaufer}, A. and {Kerber}, F., eds.,
   471 (2008).

\bibitem{Rau2011}
Rau, U. and Cornwell, T.~J., ``{A multi-scale multi-frequency deconvolution
  algorithm for synthesis imaging in radio interferometry},'' {\em \it
  \aap}~{\bf 532},  A71 (July 2011).

\bibitem{Hager2005}
Hager, W.~W. and Zhang, H., ``{A New Conjugate Gradient Method with Guaranteed
  Descent and an Efficient Line Search},'' {\em SIAM Journal on
  Optimization}~{\bf 16},  170--192 (Jan. 2005).

\bibitem{Hager2006}
Hager, W.~W. and Zhang, H., ``{A New Active Set Algorithm for Box Constrained
  Optimization},'' {\em SIAM Journal on Optimization}~{\bf 17},  526--557 (Jan.
  2006).

\bibitem{Hofmann1993}
Hofmann, K. and Weigelt, G., ``{Iterative image reconstruction from the
  bispectrum},'' {\em \aap}~{\bf 278},  328--339 (1993).

\bibitem{Lacour2008}
Lacour, S., Meimon, S., Thi\'{e}baut, E., Perrin, G., Verhoelst, T., Pedretti,
  E., Schuller, P., Mugnier, L., Monnier, J., Berger, J., and Others, ``{The
  limb-darkened Arcturus; Imaging with the IOTA/IONIC interferometer},'' {\em
  \aap}~{\bf 485},  561--570 (2008).

\end{thebibliography}
\bibliographystyle{spiebib} 

\end{document}